# An Efficient Game Theory-Based Power Control Algorithm for D2D Communication in 5G Networks

Abdu Saif[1], Kamarul Ariffin bin Noordin[1*], Kaharudin Dimyati[1*],
Nor Shahida Mohd Shah[2], Yousef Ali Al-Gumaei[3], Qazwan Abdullah[2] and Kamal Ali Alezabi[4]

[1]Department of Electrical Engineering, Faculty of Engineering, Universiti Malaya, 50603, Kuala Lumpur, Malaysia.
[2]Faculty of Engineering Technology, Universiti Tun Hussein Onn Malaysia, 84500 Bukit Pasir, Muar, Johor, Malaysia.
[3]Department of computer and information science, North Umbria University, Newcastle Upon Tyne, Tyne and Wear, UK,
[4]Institute of Computer Science and Digital Innovation, UCSI University, 56000 Cheras, Kuala Lumpur, Malaysia.

*Corresponding author: E-mail: kamarul@um.edu.my, kaharudin@um.edu.my

## *Abstract*

Device-to-Device (D2D) communication is one of the enabling technologies for 5G networks that support proximity-based service (ProSe) for wireless network communications. This paper proposes a power control algorithm based on the Nash equilibrium and game theory to eliminate the interference between the cellular user device and D2D links. This leads to reliable connectivity with minimal power consumption in wireless communication. The power control in D2D is modeled as a non-cooperative game. Each device is allowed to independently select and transmit its power to maximize (or minimize) user utility. The aim is to guide user devices to converge with the Nash equilibrium by establishing connectivity with network resources. The proposed algorithm with pricing factors is used for power consumption and reduces overall interference of D2Ds communication.

The proposed algorithm is evaluated in terms of the energy efficiency of the average power consumption, the number of D2D communication, and the number of iterations. Besides, the algorithm has a relatively fast convergence with the Nash Equilibrium rate. It guarantees that the user devices can achieve their required Quality of Service (QoS) by adjusting the residual cost coefficient and residual energy factor. Simulation results show that the power control shows a significant reduction in power consumption that has been achieved by approximately 20% compared with algorithms in [11].

## 1 Introduction

With the increasing demand for high speed and efficient data transmission, spectrum resource limitation is a challenge faced regarding this demand. Thus, Device-to-Device (D2D) communication is a promising solution to cope with resource limitations, making the network spectrum more efficient and establishing a link between devices via centralized and decentralized communication. The source and destination node's power consumption are still challenging in wireless D2D communication for achieving energy-efficient with reliable connectivity.

The power control algorithms in [1-4] resulting from the theoretical game approach are decentralized, in which each device selects its power from a power transmitting strategy through a non-cooperative scheme. The power control algorithm for D2D communications plays an essential role in minimizing power consumption while saving energy in wireless network communication.

The utility function is one of the considerations for power control distribution in the space of D2D and cellular systems. However, the amount and the unit price of jamming services and jamming power form a problem in various utility functions [5]. Nevertheless, the Stackelberg power control algorithm leads to enhances the physical layer of security underlying D2D communication and designs a technique that minimizes the interference. Consequently, it uses the Stackelberg game theory to prove the existence of the unique solution of the Nash equilibrium [6]. The existing research body on wireless communication has established a power control algorithm that is a hot topic for researchers. The recognition of power control plays a critical role in saving energy, minimizing power consumption by eliminating interference. The pre-requisite for the QoS guaranteed in the uplink communication is the power control with the pricing factor. The power control is sub-divided into two sectional variables known as the optimization problem and the non-cooperative distributed game [7].

The non-cooperative game and utility function will increase the energy network lifetime connectivity of D2D communication. In addition, the pricing factor is used to consume the power between D2D communication while extending the coverage based on the cost function and price factors. The non-cooperative game convergence has been proven with the Nash Equilibrium (NE) point for every device in the uplink transmission work in [8].

In this context, a better equilibrium solution has been achieved by adjusting the residual convergence of NE based on the residual energy factor. Then, the power control can be obtained and evaluated by the non-cooperative game scheme, resulting from the power consumption and energy efficiency [9]. The distribution of the power control problem requires an algorithm in which each terminal independently uses its transmission to choose a power level to maximize the utility of the user devices. Thus, a utility function has been designed to solve the power control problem and eliminating the D2D link and cellular user devices' interference to meet the QoS requirements[10].

The system's capacity and energy efficiency through D2D communication have been enhanced by the game theory analysis of non-cooperative power control [11].
Therefore, the power control problem is formulated as a utility function modeled by a non-cooperative game in the downlink and the uplink communications. The downlink decodes the SINR while the uplink decodes the performance of throughput[12] .

NE has become a vital technique to prove a unique solution in the non-cooperative game for power control. The non-cooperative power has been chosen based on the equilibrium points' selection mechanism. The iterative utility functions are able to generate the optimal splitting ratio of the maximized user utility within a feasible set of user devices [13].

The energy-saving and network energy lifetime link between the source and destination of D2D are inevitable for efficient connectivity in wireless communication. Furthermore, Energy Harvesting (EH) is a priming technique to cope with power consumption in the wireless network [14].

The power control strategies are considered a viable way to enhance wireless network throughput performance based on reducing user devices' consumption power and eliminating interference with other user devices. However, the utility of the user device's varied value with pricing factors acts to discharged per unit price by adjusting their power in a distributive manner. Consequently, the network adjusts and decides the unit price that maximizes the revenue in the sum of individual device payments for all the user devices during establishing connectivity [15].

According to Long Yu, it has been proven that the average utility of user devices of the Stackelberg Equilibrium (SE) is lower than that of NE since the jammer can reduce the user devices utilities by utilizing the power control advantage to improve the user utility [16].

### 1.1 Contribution

This paper aims to develop and derive a new utility function for power control in D2D communication based on a game theory. The power control algorithm is applied to the uplink of D2D communications for highly reliable connectivity yet consumes minimal power. To achieve this, the paper sets constraints for the minimum acceptable SINR within the minimum available power consumption for each device to maximize the utility function of user devices. Hence, in the case of the failure in power transmission, it will be retransmitted again to establish connectivity of the D2D link (Tx/Rx). The contributions in this paper are as follows:
- A power control algorithm has been proposed to minimize user devices' interference for efficient system communication by satisfying the SINR threshold target.
- The proposed algorithm reduces energy consumption by adjusting the residual energy factor to overcome the SINR issues and achieve a guaranteed QoS goal: lower power consumption.
- The proposed utility function with pricing promises to prolong the battery power of D2Ds and hence improve the system performance.

## 2  System Model

**Fig. 1** shows the system model with a single cell base station (BS) in a wireless network and user devices. These user devices include cellular users (UE$_1$, UE$_2$) and D2D communication (D2D$_{Tx}$-D2D$_{Rx}$) communicate with BS by using licensed and unlicensed spectrum. The power control strategy has been proposed to eliminate the interference among D2D links and the cellular system, based on the non-cooperative game theory with the pricing mechanism. Thus, $p_i^d$ denotes the power transmission for $i^{th}$ user device (UEs and D2D) while $h_i^d$ denotes the attenuation from $i^{th}$ user device to the base station. D2D communication developed connectivity with a wide range of BS coverage to boost system capacity and wireless communication efficiency. The cellular user devices are connected directly with the base station via the down and uplink channels and affect the D2D communication user device through the interference link. This work focuses on the down/uplink communication in the power control scenario that carries the data and functionality from user devices connected to the base station.

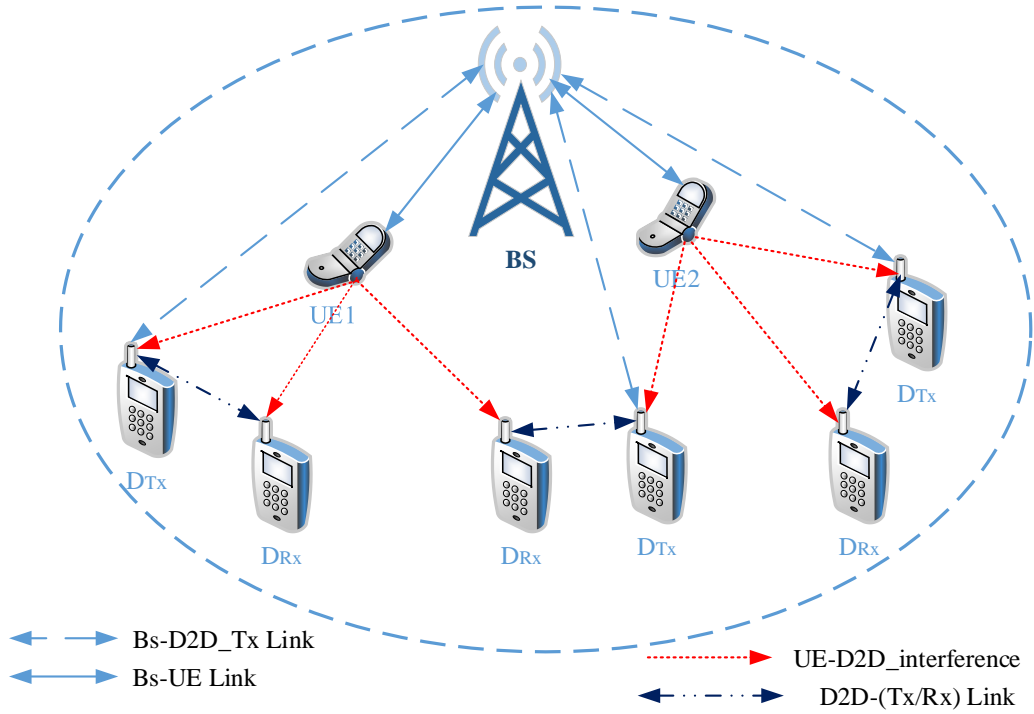

**Fig. 1.** The architecture of the system model

Hence, the purpose of developing a power control algorithm is to ensure that the SINR $\gamma_i^d$ for every user device (UEs, D2D) that remains above the threshold level for each of the $i^{th}$ user devices that denoted as follows:

$$\gamma_i^d \geq \Gamma_i^{\ d} \qquad (1)$$

Where

$\gamma_i^d$ : Denotes the $i^{th}$ user device SINR.
$\Gamma_i^d$ : Represents the $i^{th}$ user devices threshold target.
In this context, the interference of $i^{th}$ user device is expressed as a function of the power $p_i^d$ as given in the following equation:

$$I(p_i^d) = \sum_{i=1}^{N} p_i^d \quad (2)$$

Where,
$N$ : Refers to the number of user devices.
According to the game theory, three elements are considered: Players, Strategies, and Utility functions. Players represent user devices (the D2D link and cellular user), Strategies that represent the transmission power $p_i^d$ and Utility function $U_i$ that represents the payoff obtained by game players (user devices) after making the decision.

## 2.1 Utility function and NE

There are various proposals regarding the design of the utility function of user devices. However, the game theory constraint in developing user device utility based on physical output and unsatisfied game outcome [17]. Therefore, the user devices that include cellular user and D2D communications with N players and transmission power as the strategy for each player (user devices) are considered. The user devices utility function then allocates each conceivable outcome to a particular player matric number. The higher or lower attribute of a number shows whether the outcome is preferable.

The non-cooperative game formulated describes the algorithm for power control to develop a new user device utility iteration function to improve the game outcome. Consequently, the derived power control from the user devices utility function will prove the existence and convergence of NE in the algorithm [18] and [19]. This is an act of satisfaction with the convergence to occur as soon as possible and develop the new algorithm for power control. Besides, the iterative power algorithm is used to solve the NE convergence points.

We assume that the utility function of $i^{th}$ user device is $U_i(p_i^{*d}, \gamma_i^d(p_i^{*d}))$. In this context, each user's transmission power represents each player's strategy to have achieved NE and improve the utility function unilaterally. Consequently, $i$ th user devices $\in$ N-player is satisfied as:

$$U_i\left(p_i^{*d}, \gamma_i^d(p_i^{*d})\right) \leq U_i\left(p_i^d, \gamma_i^d(p_1^{*d}, p_2^{*d}, \ldots, p_{i-1}^{*d}, p_{i+1}^{*d}, \ldots, p_i^{*d})\right) \forall \; p_i^d \quad (3)$$

The proposed algorithm aims to adjust the user device's power transmission that satisfies the SINR threshold and reduces the power of user devices. Furthermore, according to [5], a utility function is presumed to be convex-shaped and assumes non-negative values to ensure the existence of a non-negative minimum. Hence, the new proposed utility function is a strategy to enhance power consumption for various values of cost coefficient ($\alpha$). Thus, the proposed utility function is expressed as:

$$U_i = \left(\frac{\Gamma_i^d}{\alpha \Gamma_i^d + 1} - \gamma_i^d\right)^2 \quad (4)$$

Where,
$\alpha$ : Cost Coefficient
$\gamma_i^d$: SINR for an $i^{th}$ user device.

The general formula for SINR is:
$$\frac{\gamma_i^d}{p_i^d} = \frac{h_i^d}{I_i^d} \Rightarrow \gamma_i^d = \frac{p_i^d h_i^d}{I_i^d} \qquad (5)$$

Where, $I_i^d$ is the interference of the effect to $i$ user devices.

The proposed power control algorithm aims to maximize the utility function derived by all the data system devices. The D2D_(Tx/Rx) will adjust its power transmitter ($p_i^d$) to maximize its utility function $U_i(p_i^d)$ for each $i^{th}$ user device. Hence, the maximum utility function will occur at a power level of the derivative of $U_i(p_i^d)$ with respect $p_i^d = 0$ and i.e. $\frac{\partial U_i}{\partial p_i^d} = 0$. Thus, from the equations (4) and (5), the utility function rewrites as:

$$U_i = \left(\frac{\Gamma_i^d}{\alpha \Gamma_i^d + 1} - \frac{p_i^d h_i^d}{I_i^d}\right)^2 \qquad (6)$$

When we take the partial derivatives of the utility function in equation (6) concerning power, $p_i^d$ and equate them to zero, the power control iteration can be obtained to achieve the minimum energy consumption as follows:

$$\frac{\partial U_i}{\partial p_i^d} = \frac{-2\left(\frac{\Gamma_i^d}{a\Gamma_i^d + 1} - \frac{p_i^d h_i^d}{I_i^d}\right)}{I_i^d} \qquad (7)$$

Hence, the necessary condition for $i^{th}$ user devices to maximize their utility to satisfy the SINR target is achieved. The optimal (minimum) value of the D2D utility function occurs when the user device's SINR is equal to the target thresholds value. The applicable method to guarantee the QoS of D2D$_s$ is balancing the power control method in which all D2D$_s$ achieve the same target SINR. The aim is to prioritize the QoS of the D2D$_s$ by ensuring that all D2D$_s$ meet the SINR target. However, to attain a higher SNR that is more significant in the target value, the D2Ds require little power in their transmission. Hence, it preserves their battery energy and network energy lifetime while minimizing cross-tier interference. To achieve the reduction in power consumption and the required SINR of D2D communication while mitigating the total interference in the D2D network through the power control game, that payoff utility function. Thus, the transmit power of $i^{th}$ user device at $(k+1)^{th}$ the time step can be expressed as follows (appendix 1):

$$p_i^{k+1} = \left(\frac{\Gamma_i^d}{a\Gamma_i^d + 1}\right)\frac{p_i^k}{\gamma_i^k} \qquad (8)$$

Equation (8) is further simplified as $\frac{\partial U_i}{\partial p_i^d} = 0$ and $p_i^d > 0$, where the $p^{(k+1)}$ is the transmission power of the $i^{th}$ D2D link at the $k^{th}$ time instant. Subsequently, each D2D link measures its current target SINR ($\gamma_i^k$) and tries to achieve its target in the next step [8].

**2.2 The utility function with a pricing factor**

The power pricing function aims to assist D2Ds by using lower transmission power based on the non-cooperative game. When using a high-power transmission, the high cost of devices employs this strategy. Thus, the pricing can be sufficient to reduce the nearest D2D$_s$, which uses low power transmission to establish the devices link. **Fig. 2** shows the performance of $i^{th}$ user device utility versus the power to transmit with/without pricing and function effects of the energy-efficient non-cooperative power control game. The user devices use lower power $P_2$ in utility pricing compared to $P_1$ in the case of utility only (without pricing). In this context, each user device adjusts its power transmission to maximize its utility price in a distributed manner. As an estimate of NE, the results balance all user devices' communications power to create a balance between the power transmission of user devices. Furthermore, assuming the pricing function can affect the utility function in equation (4). In that case, the derivative must be found to obtain the iteration power control algorithm with the pricing term factor. The algorithm includes the pricing function for power transmission propagation. Then, the user devices adjust the power level to maximize the net utility (utility pricing).

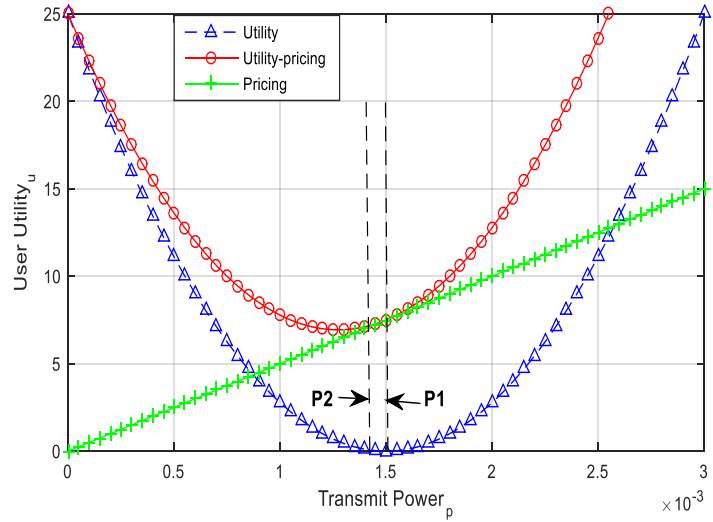

**Fig. 2.** Effect of the power pricing function on the energy-efficient non-cooperative power control game.

From equation (4), the negative pricing function $-c_i p_i^d$ is added to the utility function, and it can be formulated as follows:

$$U_i = \left(\frac{\Gamma_i^d}{\alpha\,\Gamma_i^d + 1} - \frac{p_i^d h_i}{I_i^d}\right)^2 - c_i p_i^d \qquad (9)$$

Where
$c_i$: The pricing factors.
Hence, when the partial derivatives of equation (9) for power are taken, the power control with pricing can be obtained to achieve strategies of optimum value. Then, the new power control iteration with the pricing factor can be rewritten as follows appendix (2):

$$p_i^{k+1} = \left(\frac{(ac\Gamma_i^d + 2\Gamma_i^d \frac{\gamma_i^k}{p_i^k} + c)}{2(a\Gamma_i^d + 1)}\right)\left(\frac{p_i^k}{\gamma_i^k}\right)^2 \qquad (10)$$

$$p^{k+1} = \frac{h_i^k p_i^k \Gamma_i^d}{\gamma_i^k (\alpha \Gamma_i^d + 1)} - \frac{c_i^d}{2} \qquad (11)$$

Where,
$p^{(k+1)}$ : Transmit power of $i^{\text{th}}$ user devices at $(k+1)^{th}$ time step; and
$\gamma_i^k$     : SINR of the $i^{\text{th}}$ user devices at $k^{th}$ a time step.

The utility function with pricing acts to reduce the power consumption among D2D$_s$ in the proposed power control algorithm. The following advantages are highlighted as follows:
- Increases the lifetime of the batteries of D2D$_s$ devices.
- Reduces overall interference that can harm cellular users and D2D$_s$ users for both in-band and out-band networks.
- Guarantees the QoS of the D2D communication.
- Lower interference of D2D$_s$ Ad-hoc network resulting in higher acceptance rate in the admission control of the devices.

The uplink power control challenges are the limited power transmission of user devices due to battery barriers. The low computational capability of mobiles and the near/far effects. In this regard, the pricing factor's role and the utility function to find the power iteration is brought forward.

Furthermore, all user devices will meet their SINR constraints with a lower power level achieved by the game approach to uplink D2D communication for the low-range two-tier in the proposed algorithm.

However, an efficient pricing technique would be required to handle the cross-tier interference. The iteration method used in the proposed power control algorithm was the fixed-point iterative method with slower convergence. On the other hand, some researchers proposed that to maximize benefits, the pricing factor is added to the utility to benefit selfishly, and it must be semi-concave. The ideal point will be chosen in the range of experimental parameters, such as maximum and minimum power, depending on other user devices' behavior[16].

## 3 Convergence

The new algorithm will prove convergence in the iteration of a unique NE's value for the proposed power control function. The existence of a unique value for NE has been verified.

It shows that if $p^{(k+1)} = f(p)$, a fixed point of the algorithm exists, and the function, $f()$ satisfies the following three conditions:
1. $f(p_i^d) > 0,$ shows positivity.
2. $p_i^d > p_i^{\tilde{} d} \Rightarrow f(p_i^d) > f(p_i^{\tilde{} d})$   shows monotonicity.
3. $f(\alpha p_i^d) < \alpha f(p_i^d)$ shows scalability.

Then, the algorithm has been verified the convergence to a unique and fixed point. Thus, to prove the convergence of the equation (8) has to attain the following conditions:

### 3.1 Positivity: $f(p_i^d) > 0$

From equation (8), it can be verified that $f(p_i^d) > 0$, which is achieved when:

$$f(p_i) > 0 \Rightarrow \frac{p_i^k}{\gamma_i^k}\left(\frac{\Gamma_i^d}{\alpha\Gamma_i^d + 1}\right) > 0$$

$$\text{Then, } \frac{p_i^k}{\gamma_i^k} > 0 \Rightarrow p_i^k > 0 \text{ and } \left(\frac{\Gamma_i^d}{\alpha\Gamma_i^d+1}\right) > 0 \Rightarrow \Gamma_i^d > 0 \quad (12)$$

### 3.2 Monotonicity   $p^{(k+1)} = f(p^k)$

From equation (8), In this case, if $p > \bar{p} \Rightarrow f(\bar{p}) > f(p)$, then the first-order partial derivative is set as follow:

$$\frac{\partial f(p_i^d)}{\partial(p_i^d)} = \frac{\Gamma_i^d}{\gamma_i^d(\alpha\Gamma_i^d + 1)} \quad (13)$$

For the verification of the monotonicity, the conduction to make the equation (13) positive is $\frac{\partial f(p_i^d)}{\partial p_i^d} > 0$. It is significant to have a fast response function with respect for power transmission $p_i^d$; thus:

In the case of $p > \bar{p} \Rightarrow f(\bar{p}) > f(p)$

$$\frac{p_i^d}{\gamma_i^d}\left(\frac{\Gamma_i^d}{\alpha\Gamma_i^d+1}\right) > \frac{1}{\gamma_i^d}\left(\frac{\Gamma_i^d}{\alpha\Gamma_i^d+1}\right) \Rightarrow p_i^d > 1 \quad (14)$$

Since the positivity of $0 < \alpha < 1$ has been proven, the consumption power is lesser than the SINR threshold target, this means that the iterative power function is positivity and monotonicity.

### 3.3 Scalability: $f(\alpha p) < \alpha f(p), \forall \alpha > 1$

In the final convergence stage, the scalability condition can be verified from equation (8), we obtain the following.

$$\frac{\alpha p_i^d}{\gamma_i^d}\left(\frac{\Gamma_i^d}{\alpha\Gamma_i^d + 1}\right) < \alpha\left(\frac{p_i^d}{\gamma_i^d}\left(\frac{\Gamma_i^d}{\alpha\Gamma_i^d + 1}\right)\right) \Rightarrow \frac{\alpha p_i^d}{\gamma_i^d}\left(\frac{\Gamma_i^d}{\alpha\Gamma_i^d + 1}\right) < \left(\frac{\alpha p_i^d}{\gamma_i^d}\left(\frac{\alpha\Gamma_i^d}{\alpha\Gamma_i^d + 1}\right)\right)$$

$$\Rightarrow \left(\frac{\Gamma_i^d}{\alpha\Gamma_i^d + 1}\right) < \left(\left(\frac{\alpha\Gamma_i^d}{\alpha\Gamma_i^d + 1}\right)\right) \Rightarrow \alpha\Gamma_i^{2d} + 1 < \alpha^2\Gamma_i^{2d} + \alpha\Gamma_i^d$$

$$\Rightarrow \alpha^2\Gamma_i^{2d} + \alpha\Gamma_i^d - \alpha\Gamma_i^{2d} < 1 \Rightarrow \alpha\Gamma^2(\alpha - 1) + \alpha\Gamma_i^d < 1$$

$$\Rightarrow \alpha\Gamma_i^d(\Gamma_i^d(a - 1) + 1) < 1 \Rightarrow \alpha\Gamma_i^d < \frac{1}{(\Gamma_i^d(a - 1) + 1)}$$

$$\Rightarrow \alpha > \frac{1}{\Gamma_i^d(\Gamma_i^d(a - 1) + 1)} \Rightarrow \alpha < 1$$

$$(15)$$

Therefore, the positivity of scalability has been proven. Hence, the algorithm of power control satisfies the convergence and the unique value of the NE point.

## 3.4 NE Existence

The unique solution for the Nash algorithm's algebraic equations has been proven. Then the existence of a unique solution to the proposed algorithm for the power control equation has been verified. This can be achieved using the implicit function to prove the unique solution for the power iteration function in equation (8). Besides, $-c_i p_i^d$ was added to the function to balance the power function and achieve $f(p_i^d) = 0$. Then, based on the implicit function theorem, a function is established based on equation (3) and (8), where $f_i(p_i, p_n, h_i, h_n, \alpha_i, \alpha_n)$.

$$f_i(p_i^d) = -p_i^d + \frac{p_i^d}{\gamma_i^d}\left(\frac{\Gamma_i^d}{\alpha\Gamma_i^d + 1}\right) \Rightarrow f_i(p_i^d) = -p_i^d + \frac{1}{\sum_{i=n+1}^{N} p_n h_n}\left(\frac{\Gamma_i^d}{\alpha\Gamma_i^d + 1}\right) = 0$$

Where $i = 1, 2, \ldots\ldots N$ (16)

In the implicit function theorem, if a unique NE solution exists, then the Jacobian Matrix, $\frac{\partial F_i}{\partial p_i}$ must be non-singular at the point of existence, as shown in the following matrix:

$$\frac{\partial F_i(p_i, p_n, h_i, h_n, a_i, a_n)}{\partial p_i} = \begin{vmatrix} \frac{\partial F_1}{\partial p_1} \frac{\partial F_1}{\partial P_2} \cdots \cdots \frac{\partial F_1}{\partial P_N} \\ \frac{\partial F_2}{\partial P_1} \frac{\partial F_2}{\partial P_2} \cdots \cdots \frac{\partial F_2}{\partial P_N} \\ . \\ \frac{\partial F_i}{\partial P_1} \frac{\partial F_i}{\partial p_2} \cdots \cdots \frac{\partial F_i}{\partial P_N} \end{vmatrix} \Rightarrow \frac{\partial F_i}{\partial p_i} = \begin{vmatrix} A_{11} B_{12} \cdots \cdots B_{1N} \\ 0 A_{22} \cdots \cdots B_{2N} \\ . \\ 00 \cdots \cdots \cdots A_{iN} \end{vmatrix}, \quad (17)$$

Where,

$$A_i = -1 - \frac{\Gamma_i^d}{p_i^2 h(\alpha\Gamma + 1)} \quad , \quad B_i = -\frac{\Gamma_i^d h}{(p_{i1}h + p_{i2}h)^2(\alpha\Gamma_i^d + 1)}$$

As presented above, the elements below the main diagonal line of the matrix are all zeroes. Equation (16) shows that the Jacobin Matrix is triangular. The diagonal elements determine the values of the triangle. The matrix is, therefore, a non-singular matrix. Hence, the convergence of the algorithm for the power control iteration and the existence of the power control iterative function's unique solution have been proven.

**Table 1.** Simulation Parameters

| Symbol | Parameter Name | Value |
|---|---|---|
| $p_i^d$ | User devises Transmission power | 8mW |
| $\Gamma_i^d$ | Threshold target | 5 |
| N | Number of user devices | 20 |
| $\alpha$ | Cost coefficient | (0.0,0.01,0.5) |
| $\sigma$ | Noise power | 50-15 |
| P$_{max}$ | Base Station Max transmission power | 100 mW |
| t | Number of iterations | 90, 20 |
| c | Pricing factor | 5100 |

## 4   Results and Discussion.

In this section, simulation results demonstrating the performance of the proposed schemes are presented. The results are evaluated in terms of the energy efficiency of the average power consumption, the number of D2D communication, and the number and iterations. It is assumed that user devices that included cellular user devices and D2D communications are randomly distributed in the range of base station coverage.

The simulation results are presented in **Fig. 3, Fig. 4,** and **Fig 5.** Those figures have been computed based on the changing iteration effect and the average power for $D2D_s$ communication.

The algorithm has been tested by setting the value of α = [0, 0.02] and the number of devices N = 20. The iteration ranges from 0 to 90, and the average power is from $10^{-10}$ (w) to $10^{-7}$ (w). As shown in **Fig. 3**, the number of both D2D and iteration increases from 0 to 16 and 10 to 50, respectively, with $\alpha = 0$. However, D2D increased from 16 to the highest number of 18, and the iteration risen from 50 to 90 with $\alpha =0$. Conversely, the number of D2D has been increased from 16 to 18, and the iteration decreases from 90 to 55 when $\alpha =0.02$. Hence, for any value of $\alpha$ between $0 < \alpha < 1$, the more iteration decreases, and the more significant number of D2D reaches its maximum. Therefore, this result has proved the power control algorithm's effectiveness for increasing $D2D_s$ communication based on the cost coefficient, α.

Furthermore, **Fig. 4** shows the performance of average power with various D2D pair communication when $\alpha = 0$, $a = 0.02$. The average power is increased from $10^{-10}$ to $10^{-7}$ (w), and D2D risen from 2 to 20, in the case of the cost coefficient, $\alpha =0$. However, the average decreases from $10^{-7}$ to $10^{-8}$ (w), when D2D risen from 2 to 20, in the cost coefficient, $\alpha = 0.02$. As a result, under the value range of the cost coefficient $0 < \alpha < 1$, the system's ability to allow more significant D2D connections with less power consumption has improved. Therefore, the proposed algorithm for power control significantly saves power consumption, decreases the number of iterations, and increases D2D communication in different values of the cost coefficient (α). Hence, the proposed algorithm has achieved power control in the D2D network to satisfy the target value for SINR.

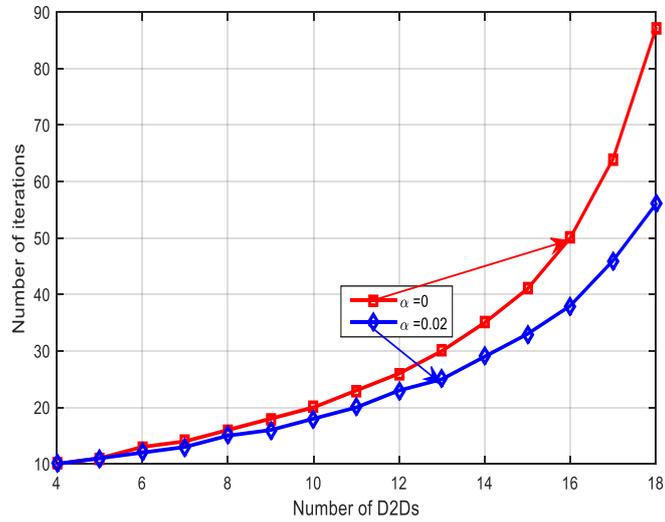

**Fig. 3.** Number of D2Ds versus iterations

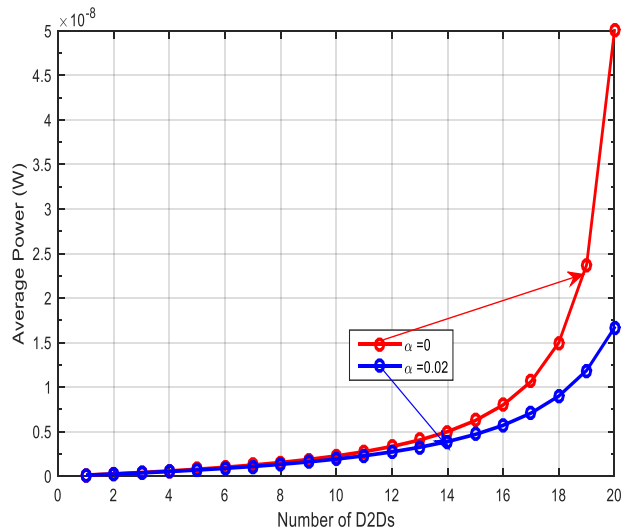

**Fig. 4.** Number of D2Ds versus average power

### 4.1 Power consumption

This section presents power consumption and the convergence rate in the proposed power control algorithm. The results have shown that higher convergence might be achieved with lesser power usage while still meeting the SINR threshold. Therefore, the essential priority of D2D communication satisfies the SINR threshold to achieve reliable and sustainable connectivity to improve the quality of the coverage. As a result, an important consideration, in this case, is the quantity of power consumption required to meet the SINR threshold. In this regard, the proposed algorithm for power control has proved that energy could be saved in the different values of $0 < \alpha < 1$. Furthermore, the proposed power control method has used less power while communicating with a more significant number of D2Ds and met the QoS requirements and the network system's data rate.

**Fig. 5** demonstrates the average SIR with iteration changes when the pricing factor equals c = 0 and c = 5100. The number of both the SINR and iteration increases from 0 to 20 and 0 to 5, respectively, with $c = 0$. However, the iteration risen from 1 to the highest number, 20, and the SINR has been increased from 0 to 5, at $c = 0$. Conversely, when the number of iterations increases from 1 to the highest number 20, the SINR decreased from 4.9 to 4.6 at c = 5100.

Hence, for any value of pricing factor ($c$) between $0 < c < 5100$, the SINR decreases by the varied significance of $c$, and the number of the iterations reaches its maximum. This result has proved that the power control algorithm's pricing factor can minimize the SINR due to decreased interference between D2D$_s$ communication. The aim to maximize the acceptable average of SINE and minimize the efficient probability of error. The power control algorithms have achieved acceptable SINR compared to related work. However, the iteration number will be convergence in a different manner.

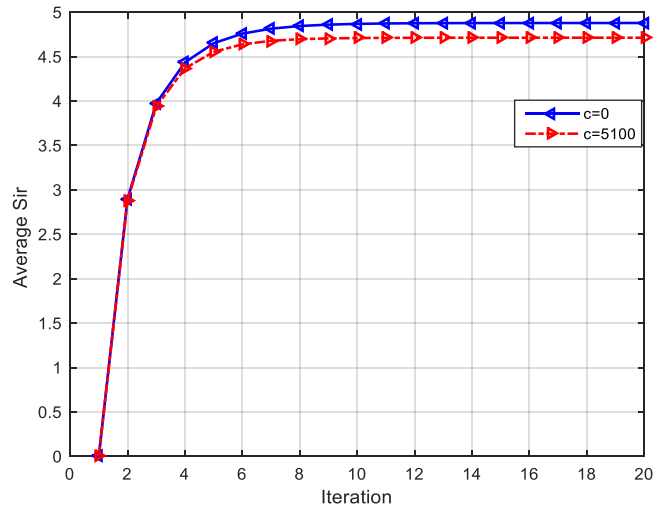

.

**Fig. 5**. Average SINR versus iteration

The comparison curves of the average transmission powers resulting from the algorithms are shown in **Fig. 6,** which illustrates that when the average power consumption of pricing factor c = 0 and c = 5100, then the algorithm has a significant reduction compared to both cases. This indicates the amount of interference measured at the D2D communication or the cellular system's interference. In this context, the proposed algorithm is the best for maximizing the spectrum's sharing and ensuring QoS in both cellular users and the D2D network.

147

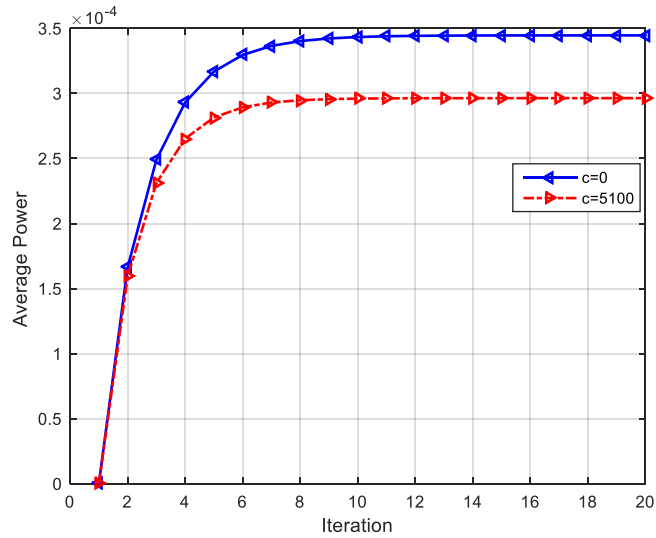

**Fig. 6.** Average power versus iteration

**Fig. 7.** shows the average power consumption of the proposed power control algorithm and compares it with other related work presented in [11] based on user devices' average power consumption and the number of iterations. It was evident from the figure that the proposed algorithm is performing better than other algorithms in average power consumption versus the number of iterations. For example, the proposed power control algorithm's average power increased with iteration from 0 to 0.2 mW. However, the other algorithm's sequences risen from 0 to (0.25,0.3,0.37,0.25) for the [11] algorithm, hyperbolic algorithm, norm2 algorithm, and CDPC algorithm, respectively.

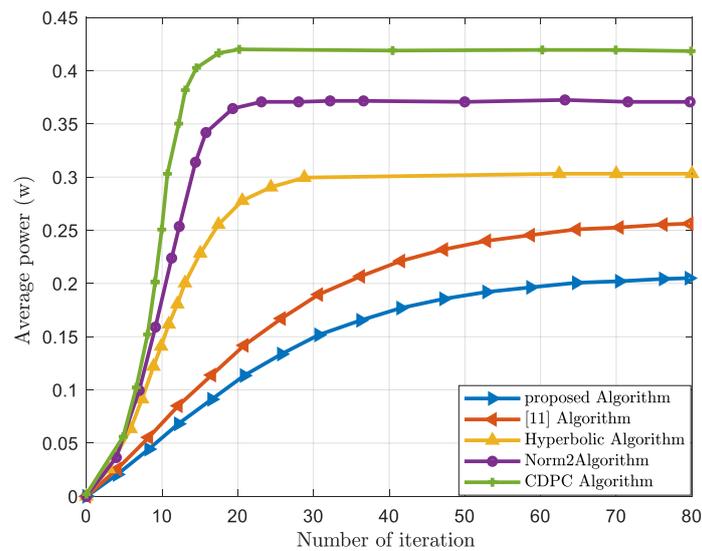

**Fig. 7.** Comparison of average power versus iteration for the proposed power control algorithms with related work in [11]

For **Table 2.** the proposed approach leads to a slower convergence rate and low power consumption with a high number of iterations. Therefore, this indicates that a trade-off between average power consumption and convergence rate is essential.

**Table 2.** Comparison convergence summary

| Algorithms | Average power (W) | Iterations |
|---|---|---|
| CDPC Algorithm | 0.43 | 20 |
| Norm2Algoritm | 0.37 | 25 |
| Hyperbolic Algorithm | 0.3 | 30 |
| Ref [11] Algorithm | 0.25 | 60 |
| Proposed Algorithm | 0.2 | 62 |

# 5   Conclusion

This paper proposed a new algorithm that has been derived from a newly developed utility function based on power control and game theory. This has produced a system that accommodates a more significant number of D2D communications with lesser power consumption. The proposed power control algorithm considers the cost coefficient and the pricing function factor. The simulation results revealed that the proposed algorithm could mitigate the SINR and establish a lower complexity link between D2D$_s$ communication.

Besides, the proposed algorithm has proved NE and satisfied the unique Nash point solution. The simulation results show that the proposed algorithm significantly saves the power consumption and increases the number of D2D$_s$ in the different values of the cost coefficient, α, within the range of $0 < \alpha < 1$. Besides, the proposed algorithm achieves a slower convergence rate and lower power consumption with a higher number of iterations compared to other algorithms.

## Appendixes

### Appendix (1)

According to equation (4), the first derivation of $U_i$ on $p_i^d$ is computed:

$$U_i = \left(\frac{\Gamma_i^d}{\alpha\Gamma_i^d+1} - \gamma_i^d\right)^2, Where\ \gamma_i^d = \frac{p_i^d h_i^d}{I_i^d} \Rightarrow U_i = \left(\frac{\Gamma_i^d}{\alpha\Gamma_i^d+1} - \frac{p_i^d h_i^d}{I_i^d}\right)^2$$

The first derivative for equation

$$\frac{\partial U_i}{\partial p_i^d} = \frac{-2\left(\frac{\Gamma_i^d}{a\Gamma_i^d + 1} - \frac{p_i^d h_i^d}{I_i^d}\right)}{I_i^d}$$

$$\frac{\partial U_i}{\partial p_i^d} = 0 \Rightarrow \frac{-2\left(\frac{\Gamma_i^d}{a\Gamma_i^d + 1} - \frac{p_i^d h_i^d}{I_i^d}\right)}{I_i^d} = 0 \quad \text{then} \quad \left(\frac{\Gamma_i^d}{a\Gamma_i^d + 1} - \frac{p_i^d h_i^d}{I_i^d}\right) = 0 \Rightarrow \frac{p_i^d h_i^d}{I_i^d} = \frac{\Gamma_i^d}{a\Gamma_i^d + 1}$$

$$p_i^d = \left(\frac{\Gamma_i^d}{a\Gamma_i^d + 1}\right)\frac{I_i^d}{h_i^d}, \quad \because \frac{I_i^d}{h_i^d} = \frac{p_i^d}{\gamma_i^d}$$

$$\text{then,} \quad p_i^d = \left(\frac{\Gamma_i^d}{a\Gamma_i^d+1}\right)\frac{p_i^d}{\gamma_i^d}$$

Thus, the transmit power of ith user device at $(k + 1)^{th}$ the time step can be expressed as follows:

$$p_i^{k+1} = \left(\frac{\Gamma_i^d}{a\Gamma_i^d+1}\right)\frac{p_i^k}{\gamma_i^k} \qquad (18)$$

## Appendix (2)

According to equation (4), the first derivation of $U_i$ on $p_i^d$ in the case of the added pricing factor, $c_i p_i^d$ is computed:

$$U_i = \left(\frac{\Gamma_i^d}{\alpha\Gamma_i^d + 1} - \gamma_i^d\right)^2 - c_i p_i^d, \text{Where } \gamma_i^d = \frac{p_i^d h_i^d}{I_i^d}$$

The negative pricing function $-c_i p_i^d$ is added to the utility function:

$$\Rightarrow U_i = \left(\frac{\Gamma_i^d}{\alpha\Gamma_i^d + 1} - \frac{p_i^d h_i^d}{I_i^d}\right)^2 - c_i p_i^d$$

The first derivative for equation (11),

$$For \; \frac{\partial U_i}{\partial p_i^d} \Rightarrow \frac{\partial U_i}{\partial p_i^d} = \frac{-2\left(\frac{\Gamma_i^d}{a\Gamma_i^d + 1} - \frac{p_i^d h_i^d}{I_i^d}\right)h_i^d}{I_i^d} - c$$

$$\Rightarrow p = \left(\frac{(ac\Gamma_i^d + 2h_i^d \Gamma_i^d + cI_i^d)I}{2(a\Gamma_i^d + 1)(h_i^d)^2}\right)$$

$$p_i^d = \left(\frac{(ac\Gamma_i^d + 2\Gamma_i^d \frac{\gamma_i^d}{p_i^d} + c)(I_i^d)^2}{2(a\Gamma_i^d + 1)(h_i^d)^2}\right)$$

$$p_i^{k+1} = \left(\frac{(ac\Gamma_i^d + 2\Gamma_i^d \frac{\gamma_i^k}{p_i^k} + c)}{2(a\Gamma_i^d + 1)}\right) \left(\frac{p_i^k}{\gamma_i^k}\right)^2$$

For $(k+1)^{th}$ iteration, the new power control iteration with the pricing factor:

$$p_i^{k+1} = \left(\frac{(ac\Gamma_i^d + 2\Gamma_i^d \frac{\gamma_i^k}{p_i^k} + c)}{2(a\Gamma_i^d + 1)}\right) \left(\frac{p_i^k}{\gamma_i^k}\right)^2 \qquad (19)$$

## Acknowledgment

The work was supported by Universiti Malaya under the EPSRC grant DARE project no. EP/P028764/1 (UM IF035A-2017 & UM IF035-2017) and Universiti Tun Hussein Onn Malaysia under the Ministry of Higher Education Malaysia Fundamental Research Grant Scheme (FRGS/1/2018/TK04/UTHM/02/14).

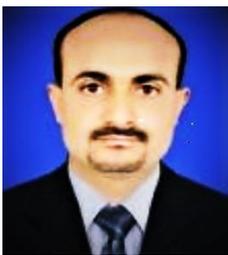
**Abdu Saif** received his Master of Science degree in Project Management in 2017 from the University of Malaya, Malaysia and a B.E. Degree in Electronics communication in 2005 from IBB University, Yemen. He has more than nine years of industrial experience in telecommunications companies. He is currently pursuing his Doctorate Degree in Electrical Engineering (major in wireless communication) from the Faculty of Engineering, University of Malaya, Kuala Lumpur, Malaysia. His research interests include Wireless networks, 3D coverage by UAV, game theory, Internet of flying things, emergency management system, and public safety communication for B5G.

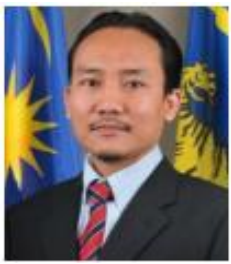
**Kamarul Ariffin** Noordin received his B.Eng. (Hons.) and M.Eng from the University of Malaya, Kuala Lumpur, Malaysia in 1998 and 2001 respectively, and his PhD in communication systems from Lancaster University in 2009, UK. He is an associate professor in the Department of Electrical Engineering, University of Malaya, Kuala Lumpur, Malaysia. His research interests mainly include resource allocation in wireless networks, cognitive, Radio networks, device-to-device communications, network modeling, and performance analysis.

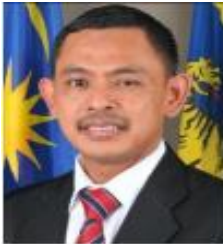

**Kaharudin Dimyati** received his Bachelor of Engineering in Electrical from the University of Malaya in 1992. He then continued his PhD in communication systems at the University of Wales Swansea, UK, in 1993 and subsequently awarded a PhD in 1996. He is currently a professor in the Department of Electrical Engineering, University of Malaya, Kuala Lumpur, Malaysia. His research interests mainly wireless communication, optical communication and coding theory. He had supervised to completion to date 20 PhD students and 33 Master by research students. He had published more than 100 papers in reputed journals. He is a member of IET (UK), IEEE (US), IEICE (Japan) and IEM (Malaysia).

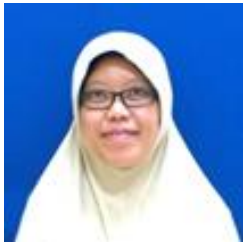

**Nor Shahida Mohd Shah** received the B.Eng. Degree from the Tokyo Institute of Technology, the M.Sc degree from the University of Malaya, and the Ph.D. degree from Osaka University, in 2000, 2003, and 2012, respectively. Since 2004, she has been with Universiti Tun Hussein Onn Malaysia. Her research interests include optical fiber devices, optical communication, antenna and propagation, and wireless communication.

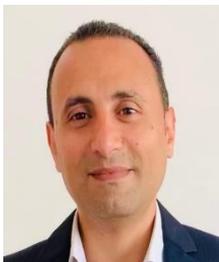

**Yousef A. Al-Gumaei** received a B.Eng degree in electrical-electronics and telecommunications engineering from IBB University, Yemen, in 2003. He received the MEngSc. And the PhD degrees in electrical-electronics and telecommunications engineering from Universiti Malaya (UM), Malaysia, in 2010 and 2017, respectively. From Jul 2017 to Jul 2019, He was an Assistant Professor in the Department of Electrical and Electronics at Al-Madinah International University (MEDIU), Malaysia. Currently, he is a visiting Research Fellow in the Faculty of Engineering and Environment at Northumbria University, Newcastle Upon Tyne, UK. From 2003 to 2007, he was a telecom engineer at the department of tele-traffic and switching at the Public Telecommunication Corporation (PTC) Yemen. In 2010, he was appointed as a lecturer in the Department of Electrical-Electronics at IBB University. His research interests include resource allocations in wireless networks, game theory, and the Internet of things, 5G.

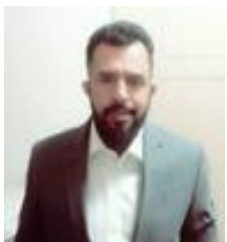

**Qazwan Abdullah** was born in Taiz, Yemen and received his Bachelor's Degree in Electrical and Electronic Engineering from Universiti Tun Hussein Onn Malaysia (UTHM) in 2013. He also received his Master's Degree in Electrical and Electronic Engineering in 2015 from the same university. Currently, he is a PhD student with research interests in control systems, wireless technology and antenna filter design.

+

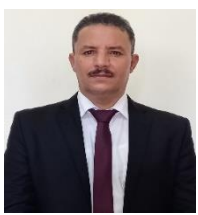

**Asst. Prof. Dr Kamal Ali Alezabi** is a senior lecturer at the Institute of Computer Science and Digital Information, UCSI University, Malaysia. He is currently the Head of Research and Postgraduate Studies, UCSI University. He was awarded his PhD from the Computer and Communication Systems Engineering Department at University Putra Malaysia (UPM) in 2017. Dr Kamal serves as Program Committee and Scientific Committee member, the International Conference on Emerging Technologies and Intelligent Systems (ICETIS 2021). Program Chair Committee member, International Conference on Emerging Technology in Computing, Communication and Electronics (ETCCE 2020). Industry Panel Assessors member, UTM.